\documentclass[aps,prl,twocolumn,groupedaddress,showpacs]{revtex4}
\usepackage{graphicx}
\begin{document}

\title{Disorder and the Supersolid State of Solid $^4$He}

\author{Ann Sophie C. Rittner and John D. Reppy}
\email{jdr13@cornell.edu}

\affiliation{Laboratory of Atomic and Solid State Physics and the
Cornell Center for Materials Research, Cornell University, Ithaca,
New York 14853-2501}

\date{\today}

\begin{abstract}
We report torsional oscillator supersolid studies of highly
disordered samples of solid $^4$He. In an attempt to approach the
amorphous or glassy state of the solid, we prepare our samples by
rapid freezing from the normal phase of liquid $^4$He. Less than two
minutes is required for the entire process of freezing and the
subsequent cooling of the sample to below 1 K. The supersolid
signals observed for such samples are remarkably large, exceeding 20
\% of the entire solid helium moment of inertia. These results,
taken with the finding that the magnitude of the small supersolid
signals observed in our earlier experiments can be reduced to an
unobservable level by annealing, strongly suggest that the
supersolid state exists for the disordered or glassy state of helium
and is absent in high quality crystals of solid $^4$He.
\end{abstract}


\maketitle

Following the discovery by Kim and Chan (KC) \cite{1,2} of the
supersolid or nonclassical rotational inertia (NCRI) state of bulk
solid $^4$He, several independent groups using the same torsional
oscillator technique have confirmed the KC supersolid results. These
include the Japanese groups of Shirahama et al.\ \cite{shirahama},
working at Keio University, and Kubota et al.\ \cite{kubota} at the
ISSP, as well as our group \cite{us} at Cornell University. In these
early experiments, the solid samples were formed by the blocked
capillary technique. In this method the fill line to the cell is
first allowed to freeze ensuring that solidification in the cell
occurs under a condition of constant average density. This technique
is known to produce relatively disordered polycrystalline samples.
The signals observed in the early experiments were small
representing, at most, a few percent of the total solid helium mass.
The Cornell experiments \cite{us} also demonstrated that the
supersolid signal could be substantially reduced through annealing
of the sample. In some cases the annealing process appeared to
eliminate the supersolid signal; i.e., it reduced the supersolid
fraction below the 0.05\% level of experimental detection. This
signal reduction upon annealing strongly suggested that sample
disorder plays an important role in supersolid phenomena. This
inference is supported by more recent work by Chan's group
\cite{mosesprl} where high quality crystals were grown under
conditions of constant pressure. The supersolid signals observed for
these constant pressure samples were somewhat smaller in magnitude
than those obtained for more disordered samples created in the same
cell by the blocked capillary method.

Recently, there has been a growing consensus in the theoretical
community \cite{ceperley,svistunov, svistunov2,dirty} that an ideal
hcp helium crystal will not exhibit the supersolid phenomenon, but
rather, some form of disorder such as vacancies, interstitials,
superfluid grain boundaries \cite{balibar,troyer}, or perhaps a
glassy or superglass phase \cite{glass,svistunov3} is required for
the existence of the supersolid state. A summary of the current
theoretical literature is given in a recent review \cite{prokofev}.

In the measurements reported here, we investigate the role of
disorder in supersolid phenomena. In order to maximize sample
disorder we have confined the solid within a narrow annular region.
The small volume and large surface area to volume ratio (S/V)
provided by this geometry allow rapid freezing and subsequent
cooling of the sample to low temperatures, ensuring a high degree of
frozen-in disorder. To provide a contrast to the disordered annular
samples, we have also studied the supersolid phenomena in an open
cylindrical geometry where slow freezing and cooling are employed to
promote the growth of large helium crystals with a relatively low
level of disorder.

\begin{figure}
\setlength{\unitlength}{1.0in}
\begin{picture}(2.0,2.5)
\put(0,0) { \makebox(1.5, 2.5)[t] {
\includegraphics[width=0.5\textwidth]{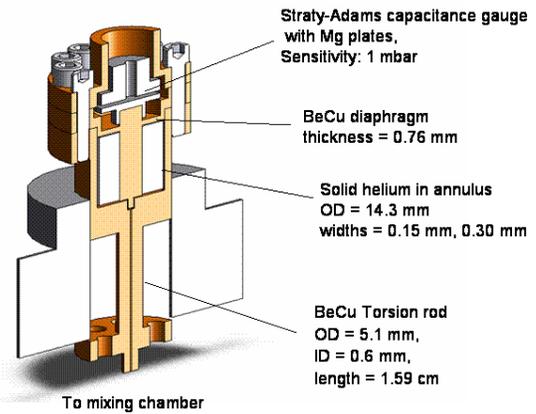}}}
\end{picture}
\caption{\label{1oscillator}Torsional oscillator: The motion of the
torsion bob is excited and detected electrostatically. A
Straty-Adams capacitance gauge, a heater, and a thermometer are
attached on top of the torsion bob. At 4 K, the resonance frequency
with the biggest insert is 874 Hz, and the mechanical quality
factor, Q = 5 $\times$ $10^5$.}
\end{figure}

The torsional oscillator cell employed for these measurements is
shown in Fig.\ \ref{1oscillator}. A special feature is a removable
top plate sealed to the body of the oscillator with a lead o-ring.
This feature allows us to vary the internal geometry of the cell
between runs. A capacitance pressure gauge, along with a thermometer
and heater, is incorporated into the top structure of the
oscillator. Cylindrical magnesium inserts can be mounted inside the
oscillator to provide a range of annular spacings. The sample volume
of the cell with the insert structure removed is 1.8 cm$^3$ with a
surface area to volume ratio, S/V = 4.6 cm$^{-1}$. With a magnesium
insert in place, the open volume of the cell is much reduced and the
surface area in contact with the solid helium is nearly doubled. At
this time we have investigated samples formed with two different
annular gaps: 0.30 mm, and 0.15 mm. For the cell with the smallest
gap, S/V = 131 cm$^{-1}$.

We determined the moment of inertia of the cell to be 51 gcm$^2$
based on the change in frequency produced by a calibrated variation
in the cell moment of inertia. We obtain the solid $^4$He moment of
inertia, $I_s$, by a calculation based on the density of the solid
helium at the melting temperature and the measured geometry of the
cell interior. In the case of the cell with the 0.15 mm gap, $I_s =
9.8 \times 10^{-3}$ gcm$^2$ for an assumed a $^4$He molar volume of
19.5 cm$^3$. The addition of this moment of inertia to the
oscillator, would produce a increase in the oscillator period of
magnitude, $\Delta P_0$ = 109 $\pm$ 4 ns. The uncertainty in $\Delta
P_0$ arises almost entirely from the measurement error in
determining the dimension of the 0.15 mm gap.

In previous torsional oscillator supersolid experiments, the
quantity, $\Delta P_0$, is obtained from the period shift seen upon
freezing. In our case the period shift that occurs during freezing
is complicated by a shift of opposite sign caused by the drop in
pressure that occurs during solid formation. The pressure
sensitivity of the oscillator period is discussed below. In the case
of our 0.15 mm gap samples the period shift due to the pressure
effect is on the order of 20 - 30\% of $\Delta P_0$. Therefore, we
have chosen to calculate $\Delta P_0$ rather than rely on a pressure
correction to the period shift seen on freezing.

Otherwise, the experimental procedures followed in these
measurements are similar to those employed in our earlier work
\cite{us}. As a first step, an empty cell run is made to determine
the temperature-dependent backgrounds for the torsional oscillator
period, dissipation, and the capacitance pressure gauge. The helium
used for our samples is commercial grade helium with a reported
$^3$He concentration of 0.3 ppm. After the cell is filled with
liquid $^4$He, the capacitance pressure gauge is calibrated at 4 K
for pressures up to 70 bar. As the pressure is raised during this
calibration, the oscillator period is observed to increase linearly
with pressure, with a sensitivity of 1.14 ns/bar. In our
measurements discussed in this letter, the samples were all formed
at pressures above hcp-bcc triple point to avoid complications due
to the bcc phase.

\begin{figure}
\includegraphics[width=0.5\textwidth]{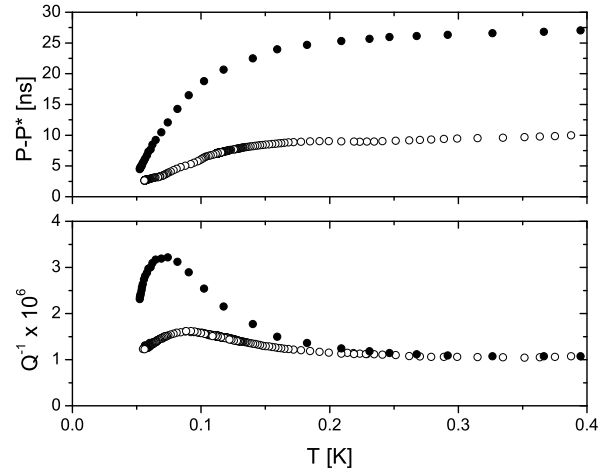}
\caption{\label{2periodanddissipation}Period and dissipation are
shown as a function of temperature for the annular cell with 0.15 mm
width. The first sample (open circles) was cooled to below 1 K over
a period of three hours. Following this low temperature run, the
sample was rapidly melted and quench-cooled (solid circles) to below
1 K in 90 s. The sample pressures were 41 bar (open circles) and 51
bar (solid circles) and the rim velocities at 300 mK were 9.7
$\mu$m/s (open circles) and 5.9 $\mu$m/s (solid circles)
respectively. The value of P* for both samples is 1.144785 ms.}
\end{figure}

In Fig.\ \ref{2periodanddissipation} we show period and dissipation,
$Q^{-1}$, data as a function of temperature for two runs with the
0.15 mm gap cell. The lower set of period data was obtained for the
first run following the initial freezing of the sample. In this case
the freezing and subsequent cooling to below 1 K took place over a
three-hour period. During this relatively slow cooling process we
believe that a certain amount of sample annealing can take place,
resulting in a relatively small supersolid signal obtained by taking
the difference between the period data and a linear fit to period
data above the supersolid transition. A further decrease in cooling
rate, taking 14 hours to cool below 1 K, did not result in an
additional decrease in signal size. The upper data set was obtained
after a "quench" cool of the sample. In this procedure, the sample
is melted by a heat pulse applied to either the cell heater or the
cell thermometer. As soon as melting has occurred, as indicated by
the pressure and period signals, the heat is turned off and the
sample rapidly freezes and cools to a temperature below 1K in a time
interval of approximately 90 seconds.

The most important feature of these quench-cooled data is the large
increase in the magnitude of the supersolid signal. For this sample,
the period reduction, $\Delta P(T)$, at 50 mK is about 22 ns and
corresponds to a supersolid fraction, $\rho_s/$$\rho\equiv\Delta
P(T)$/$\Delta P_0$, amounting to a 20\% fraction of the solid helium
moment of inertia opposed to 6 \% for the slowly frozen sample.
Similarly, the supersolid fraction in the 0.3 mm annulus varied from
4 \% (for slowly frozen samples) to 6 \% after a quench cool.
Although this signal is more than an order of magnitude larger than
any reported for previous experiments
\cite{1,2,shirahama,kubota,us}, the general temperature dependence
of both the supersolid signal and the dissipation data are similar
to that of the earlier data.

In a significant recent experiment, Sasaki et al.\ \cite{balibar}
have observed grain-boundary mediated superflow in solid samples in
contact with the superfluid phase. These authors suggest that
superflow along the surface of grain boundaries may be a possible
explanation for the supersolid signals observed by KC. Although the
small signals observed in the early experiments might be explained
by this mechanism, it is difficult imagine grain boundaries
occupying 20\% or more of the sample volume, as would be required to
explain the signals in the present experiments. A solid helium
sample with such a high concentration of grain boundaries might
better be described as a glass.

Another feature of the data shown in Fig.\
\ref{2periodanddissipation}, is an increase in the period of the
oscillator following the quench-cool. This increase in the period is
principally associated with the substantial increase in the sample
pressure of 9.7 bar resulting from the quench-cool process. We
believe that this increase is a result of inflationary pressure
arising from the increased disorder in the sample. Vacancies,
microscopic voids, or a possible glassy phase are the most obvious
candidates for the source for the inflationary pressure. If so, the
excess pressure provides a convenient measure of the relative degree
of disorder in the sample.

\begin{figure}
\includegraphics[width=0.5\textwidth]{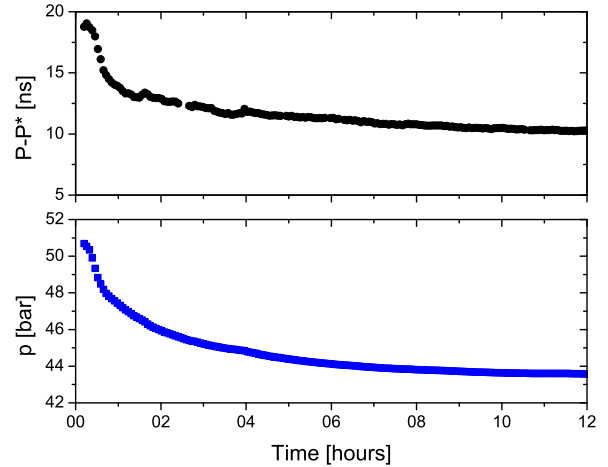}
\caption{\label{3annealing}Pressure and period are shown as a
function of time during annealing. The temperature is set to 1.57 K,
the melting temperature of the sample is 2.35 K.}
\end{figure}

Annealing of this sample at temperatures near the melting value will
lead to a reduction in the level of disorder and should also lead to
a reduction in the sample pressure. To test this idea, we have
raised the temperature of the sample to 1.57 K where thermally
activated annealing is expected on the basis of our earlier work
\cite{us}. In Fig.\ \ref{3annealing} we plot both the oscillator
period and the sample pressure as functions of time during the
annealing process. Both the pressure and the period are seen to
relax with a time constant of approximately one hour. After waiting
10 hours the pressure approaches a constant value, 7 bar below the
pre-annealed value. A further increase in the temperature to 2.0 K
produces a further drop in pressure with a much faster relaxation
time. The reduction in pressure during the annealing of disordered
solid $^4$He samples is not a new phenomenon, but has been reported
a number of times by other experimenters, most recently by
\cite{tikhii} in a paper reporting evidence for a glassy phase in
$^4$He samples grown by the blocked capillary method.

We have also studied the influence of velocity on the magnitude of
the supersolid signal for the 0.15 and 0.30 mm annular cells. In
general, the results are similar to those reported by KC for their
0.63 mm annular cell; however, we find a critical velocity which is
a factor of 4 higher than for the KC data.

\begin{figure}
\includegraphics[width=0.5\textwidth]{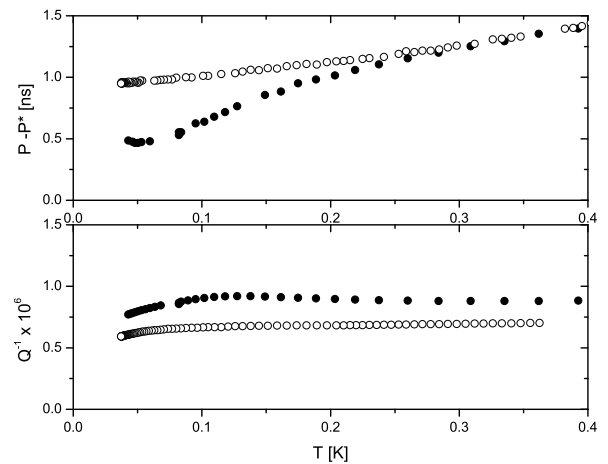}
\caption{\label{3periodanddissipationcylinder}Period and dissipation
are shown as a function of temperature in a cylindrical cell with a
volume of 1.8 $cm^3$. The sample (solid circles) was frozen and
cooled below 1 K in 2 hours. The sample pressure was 32 bar and the
rim velocity at 300 mK was 24.7 $\mu$m/s. At a velocity of 9
$\mu$m/s, the graphs are identical. We also display the empty cell
period and dissipation (open circles). The values of P* are 1.054064
ms (solid circles) and 1.052100 ms (empty cell).}
\end{figure}

To further investigate the influence of disorder on the magnitude of
the supersolid signal, we have formed solid samples in a 951 Hz
oscillator with a relatively large, 1.8 cm$^3$, open cylindrical
volume and S/V of 5.84 cm$^{-1}$. The samples formed in this cell
require two hours for the freezing of the sample and the subsequent
cooling to below 1 K. Given the relatively slow cooling through the
annealing temperature range, T $>$ 1 K, we expect a polycrystalline
sample with relatively large crystals and a significantly reduced
level of disorder as compared to the quench-cooled samples.

In Fig.\ \ref{3periodanddissipationcylinder} we display period and
dissipation signals for these samples. The supersolid signals are
very small, on the order of $3.1  \times 10^{-4}$ of the total solid
helium moment of inertia for velocities as low as 9 $\mu$m/s. Thus,
an alteration in sample preparation and the surface to volume ratio
can lead to a reduction in the amplitude of the supersolid signal of
three orders of magnitude. We believe that the extremely small
signals observed for these large open volume samples can most likely
be attributed to remnant vestigial disorder in the sample and do not
represent supersolid flow within the helium crystallites themselves.

\begin{figure}
\includegraphics[width=0.5\textwidth]{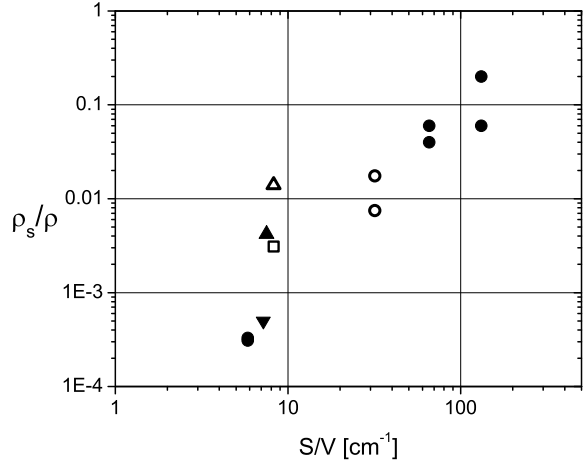}
\caption{\label{5geometry} Overview for a range of supersolid
fractions as a function of surface to volume ratios from different
researchers. All the data obtained by blocked capillary method is
above 31 bar in order to avoid complications by the bcc phase.
Cornell (closed circles); Penn State, annular cell (open circles)
\cite{2,mosesprl}, cylindrical cell T$_{melting}$ = 2.17 K (open
triangle) \cite{mosesprivate}, and cylindrical cell constant
pressure growth 26 bar (open square) \cite{mosesprivate}; Shirahama
et al. 41 bar (closed triangle) \cite{shirahama}. An upper bound on
the size of the supersolid signal set by our earlier measurements in
a square cell at 32 bar \cite{us} is displayed as an inverted closed
triangle.}
\end{figure}

In Fig.\ \ref{5geometry} we give an overview of the relation between
the surface to volume ratio and the maximum and minimum observed
supersolid signals. The data shown in this plot demonstrate a clear
trend for increasing supersolid signal with increasing surface to
volume ratio. As S/V increases, the possibility of greater frozen-in
disorder increases, since the samples can be frozen and cooled more
quickly and also because the disorder may be stabilized by a more
confining geometry.

At this time, it is still unclear what the exact mechanism for the
supersolid phenomenon may be. It is clear, however, that disorder
plays a key role in the phenomenon: increasing disorder leads to
larger supersolid signals while reducing disorder by growing higher
quality crystals has the effect of reducing the supersolid signal,
in some cases below the detectable level. An unresolved puzzle is
presented by the results of KC in vycor \cite{1} and porous gold
\cite{mosesjltp}; in both cases the S/V ratios are orders of
magnitude larger than for our cells, yet the supersolid fractions
are on the order of 2\%.

\begin{acknowledgments}
We would like to acknowledge useful conversations with M.H.W. Chan,
A.C. Clark, E. Mueller, K. Hazzard, V. Elser and G.V. Chester, and
we also thank J.V. Reppy for editorial assistance. The work reported
here has been supported by Cornell University, the National Science
Foundation under Grant DMR-060584 and through the Cornell Center for
Materials Research under Grant DMR-0520404.
\end{acknowledgments}
\bibliography{supersolid2}
\end{document}